\documentclass{ws-ijmpb}

\begin{document}

%
\catchline{}{}{}{}{}
%

\title{Intermediate Symmetries In Electronic Systems:
Dimensional Reduction, Order Out Of Disorder, Dualities, 
And Fractionalization}

\author{ZOHAR NUSSINOV}

\address{Department of Physics, Washington University, St. Louis\\
MO 63160, USA \\
zohar@wuphys.wustl.edu}

\author{CRISTIAN D. BATISTA}

\address{Theoretical Division, Los Alamos National Lab,\\
Los Alamos, NM 87545, USA\\
cdb@lanl.gov}

\author{EDUARDO FRADKIN}

\address{Department of Physics, University of Illinois,\\
1110 W. Green St., Urbana, IL 61801, USA\\
efradkin@uiuc.edu}

\maketitle

\begin{history}
\received{Day Month Year}
\revised{Day Month Year}
\end{history}

\begin{abstract}
We discuss symmetries intermediate between global and local and 
formalize the notion of dimensional reduction adduced from such symmetries.
We apply this generalization to several systems 
including liquid crystalline phases of Quantum Hall systems, 
transition metal orbital systems, frustrated spin systems, 
(p+ip) superconducting arrays, and sliding Luttinger liquids.
By considering space-time reflection symmetries,
we illustrate that several of these systems are dual to 
each other. In some systems exhibiting these symmetries, low temperature 
local orders emerge by an "order out of disorder" effect while in
other systems, the dimensional reduction precludes standard
orders yet allows for multiparticle orders (including those
of a topological nature).

\end{abstract}


\section{Introduction and Main Results}

Orders are often very loosely classified into two types:

({\bf i}) {\em Global symmetry breaking orders}. In many systems  
(e.g. ferromagnets), there is an invariance 
of the basic interactions with 
respect to global symmetry operations 
(e.g.rotations in the case of ferromagnets) 
simultaneously performed on all of the 
fields in the system. At sufficiently low temperatures
(or high coupling), such symmetries may often be ``spontaneously''
broken.

({\bf ii}){\em  ``Topological orders''}\cite{wenbook}. In some cases,
even if global symmetry breaking cannot occur, 
the system still exhibits a robust
order of a topolgical type. 
This order may only be detected by non-local correlation functions.
The most prominent examples of this 
order are afforded by gauge theories which
display local gauge symmetries.

In this article, we investigate 
{\em intermediate} (or sliding) symmetries 
which, generally, lie midway between the global symmetries
and local gauge symmetries extremes and 
provide conditions under which
local orders cannot appear. We will 
review:

({\bf 1}) a theorem \cite{BN} dictating that in many systems displaying 
intermediate symmetries, local
orders are impossible. This theorem
also gives upper bounds 
on multi-particle correlators
and suggests for 
fractionalization
in certain instances. 

({\bf 2}) how symmetry
allowed orders in such highly degenerate systems
may be stabilized by entropic fluctuations.
(This is often refered to as the ``order
out of disorder'' mechanism.) 
In classical ({\em large S}) renditions of quantum
orbital systems, orbital order is stabilized 
by thermally driven entropic fluctuations.\cite{NBCv} \cite{BCN} 
These classical tendencies may be fortified by the 
incorporation of zero point quantum fluctuations. 

({\bf 3}) a duality betwen two prominent 
systems exhibiting intermediate symmetries. 
A route by which 
this duality may be established sheds 
light on some dualities as direct consequences 
of geometrical $(Z_{2}$)
reflections in space-time. \cite{NF}

The results reported here appeared,
in full detail, elsewhere. Our aim is
to give a flavor of these
which the reader may then peruse in 
detail. For pedagogical 
purposes, we review in section (2), 
old results conerning lattice gauge theories. 
In sections thereafter, we discuss new results and constructs concerning
systems with non-local intermediate (or sliding) symmetries where 
some analogies and extensions may be drawn vis a vis 
the physics of gauge theories which 
display stronger local symmetries.

\section{What are gauge theories and local 
gauge symmetries?}
\label{what?}

 Throughout this work, we will 
consider theories defined on a lattice ${\Lambda}$. 
We start with a review of a very well known topic.
In matter coupled gauge theories, \cite{Kogut} \cite{Fradkin} 
\cite{Wegner} \cite{ZN} matter fields $(\{\sigma_{i}\}$)
reside at sites $i$ while gauge fields $U_{ij}$ lie on
links between sites $i$ and $j$.
The ${\mathbb Z}_{2}$ matter coupled to ${\mathbb Z}_{2}$ gauge field 
theory is the simplest such theory. 
On a hypercubic lattice, its
action is
\begin{eqnarray}
S = - \beta \sum_{\langle i j \rangle} \sigma_{i} U_{ij} \sigma_{j} 
- K \sum_{\Box} UUUU.
\label{intro}
\end{eqnarray} 
The first sum is over all nearest neighbor links $\langle i j \rangle$ 
while the second is the product
of the four gauge fields $U_{ij} U_{jk} U_{kl} U_{ki}$
over each minimal plaquette (square) of the lattice.
Both matter ($\sigma_{i}$) and gauge ($U_{ij}$) 
fields are Ising variables within this
theory: $\sigma_{i} = \pm 1$, $U_{ij} = \pm 1$. 
The action $S$ is invariant
under local ${\mathbb Z}_{2}$ gauge transformations
\begin{eqnarray}
\sigma_{i} \to \eta_{i} \sigma_{i}, ~  U_{ij} \to \eta_{i} U_{ij} \eta_{j} ~~~
(\eta_{i} = \pm 1)
\label{gt}
\end{eqnarray}
There is a theorem,
known as Elitzur's theorem, \cite{Elitzur} which disallows 
quantities not invariant under such these local
symmetries (e.g. $U,\sigma$)
to attain finite expectation values, $\langle \sigma_{i} \rangle
= \langle U_{ij} \rangle =0$.
In a pure gauge theory having no matter coupling
($\beta=0$ in Eq.(\ref{intro})), the only symmetry invariant
quantities are non-local quantities of a topological nature- the 
products of gauge fields 
around closed loops- the ``Wilson loops''
$W \equiv \langle U_{ij} U_{jk} ... U_{pi} \rangle$. \cite{Kogut} 
In a pure gauge theory, the asymptotic
scaling of $W$ 
with the loop size dictates what phase we are in. \cite{Kogut} 
In the presence of matter, we also find symmetry invariant
open string correlators (e.g., $\langle \sigma_{i} U_{ij} U_{jk}... 
U_{pm} \sigma_{m} \rangle$). \cite{Kogut} \cite{Fradkin} \cite{Wegner}
\cite{ZN} 
The simplest realization of Eq.(\ref{intro}) (that in two dimensions) 
exhibits a non-local (percolation) crossover as a function of 
$\beta$ and $K$. \cite{ZN} 

Electromagnetism is a gauge theory with a local $U(1)$ invariance,
\begin{eqnarray}
\sigma_{i} \to \eta^{*}_{i} \sigma_{i}, ~  U_{ij} \to \eta_{i}^{*} 
U_{ij} \eta_{j}, ~~ (\eta_{i} = e^{i \theta_{i}}, ~ \theta_{i} \in \Re).
\label{gt1}
\end{eqnarray}
Here, we set $U_{ij} = \exp[i A_{ij}]$, 
with $A_{ij} = \int_{i}^{j} \vec{A} \cdot \vec{dr}$, where $\vec{A}$ 
is the vector potential. With the complex $U(1)$ fields $U_{ij}$
and $\sigma_{1i}$, Eq.(\ref{intro}) 
is changed by the addition of
a complex conjugate.
Here, the plaquette term reads 
$[- \frac{K}{2} (U_{i} U_{jk} U_{kl} U_{li} + c.c.)]  = [- K 
\cos \Phi_{\Box}]$
with $\Phi_{\Box} = A_{ij} + A_{jk} + A_{kl} + A_{li}$ the flux
piercing the plaquette. Expanding, in the continuum limit, 
$(-K \sum_{\Box} \cos \Phi_{\Box}) 
\to (\frac{K}{2} \int dV (\nabla \times \vec{A})^{2})$, and
familiar continuum electromagnetism appears. Similarly, the first term
of Eq.(\ref{intro}) (now $-\frac{\beta}{2} \sum_{\langle i j \rangle}
(\sigma_{i}^{*} U_{ij} \sigma_{j} + c.c.)$) 
becomes, in the continuum, the standard minimal coupling term
between charged matter and electromagnetic fields. Here, the Wilson loop 
becomes the Aharonov-Bohm phase. \cite{AB} 
Higher order groups ($U(1) \times SU(2), SU(3)$)
describe the electroweak and strong interactions.

\section{What are intermediate symmetries?}

  An intermediate $d$-dimensional symmetry of a theory \cite{BN}
characterized by a Hamiltonian $H$ (or action $S$) 
is a group of symmetry transformations such that the minimal non-empty 
set of fields $\phi_{\bf i}$ changed by the group 
operations occupies a $d$-dimensional subset 
${\cal C} \subset \Lambda$. The index ${\bf i}$ denotes the sites 
of the lattice ${\Lambda}$. For instance, 
if a spin theory is invariant under flipping each individual spin then the 
corresponding gauge symmetry will be zero-dimensional 
or local. Of course flipping a chain of spins is also a symmetry, 
but the chain is not the minimal non-trivial subset 
of spins that can be flipped. In general,
these transformations can be expressed as:
\begin{equation}
{\bf U}_{lk} = \prod_{{\bf i} \in {\cal C}_l} {\bf g}_{{\bf i}k},
\label{tran}
\end{equation}
where ${\cal C}_l$ denotes the subregion $l$, ${\cal C}_l \subset \Lambda$,
and  
${\Lambda}= \bigcup_{l} {\cal C}_l$.
To make contact with known cases,
the local gauge symmetries of Eqs.(\ref{gt}, \ref{gt1})
correspond to $d=0$ as the region where 
the local gauge symmetries operate is 
of dimension $d=0$. Similarly, e.g. 
in a nearest neighbor ferromagnet
on a $D-$ dimensional lattice, 
$H = -J \sum_{\langle i j \rangle} \vec{S}_{i} \cdot \vec{S}_{j}$,
the system is invariant under a global rotation of
all spins. As the volume influenced by the 
symmetry operation occupies a $D-$ dimensional 
region, we have that $d=D$.

\section{Examples of intermediate symmetries}

{\bf{a}}) {\it Orbitals}- In transition metal (TM) systems
on cubic lattices, each TM
atom is surrounded by 
an octahedral cage of oxygens. Crystal fields
lift the degeneracy of the five 3d orbitals
of the TM
to two higher energy $e_{g}$ 
levels ($|d_{3z^{2}-r^{2}} \rangle$ and $| d_{x^{2}-y^{2}} \rangle$)
and to three lower energy $t_{2g}$ levels
($| d_{xy} \rangle$, $| d_{xz} \rangle$,
and $| d_{yz} \rangle$). 
A super-exchange calculation leads to the 
Kugel-Khomskii Hamiltonian  \cite{KK} \cite{Brink03}
\begin{equation}
\label{1}
H = \sum_{\langle r,r' \rangle} H_{orb}^{r,r'} 
(\vec{S}_{r}\cdot \vec{S}_{r'} + \frac{1}{4}).
\end{equation}
Here, $\vec{s}_{r}$ denotes the spin of the electron at site~$r$ 
and $H_{orb}^{r,r'}$ are operators acting on the orbital degrees of freedom.
For TM-atoms arranged in a cubic lattice,
\begin{eqnarray}
\label{orbHam}
H_{orb}^{r,r'} = J(4\hat{\pi}_{r}^\alpha \hat{\pi}_{r'}^\alpha 
-2\hat{\pi}_{r}^\alpha - 2\hat{\pi}_{r'}^\alpha+1),
\end{eqnarray}
where ~$\hat{\pi}_{r}^\alpha$ are orbital pseudospins
and~$\alpha=x,y,z$ is the direction of the bond~$\langle r, r' \rangle$. 

\begin{figure}[t]
\refstepcounter{obrazek}
\label{figrubick}
\centerline{\includegraphics[width=4.3in]{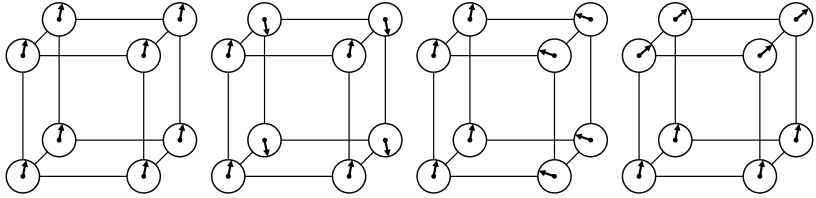}}
\medskip
\caption{From Refs$^{3~4}$. The symmetries of Eq.(\ref{symorb}) applied on
the uniform state (at left).}
\end{figure}

{\bf(i)}  In the ~$e_{g}$ compounds,
\begin{equation}
\hat\pi_{r}^x=\frac{1}{4}(-\sigma_{r}^z+\sqrt{3}\sigma_{r}^x),\qquad
\hat\pi_{r}^y=\frac{1}{4}(-\sigma_{r}^z-\sqrt{3}\sigma_{r}^x), ~~
\hat\pi_{r}^z= \frac{1}{2}\sigma_{r}^z.
\label{120_op}
\end{equation}
This also defines the orbital only ``120$^\circ$-Hamiltonian'' given by 
\begin{equation}
H_{orb} = J \sum_{r,r'} \sum_{\alpha=x,y,z} \hat{\pi}_{r}^{\alpha} 
\hat{\pi}_{r+ \hat{e}_{\alpha}}^{\alpha}. 
\label{orb}
\end{equation}
Jahn-Teller effects in $e_{g}$ compounds also lead, on their own, to
orbital interactions of the 120$^\circ$-type. \cite{Brink03} The 
``$120^{\circ}$ model'' model of Eqs.(\ref{120_op},\ref{orb})
displays discrete ($d=2$) 
$[Z_{2}]^{3L}$ gauge-like symmetries 
(corresponding to planar Rubick's cube like reflections
about internal spin directions- Fig.(\ref{figrubick})). 
The symmetry operators $O^{\alpha}$ are \cite{BN} \cite{NBCv} \cite{BCN}
\begin{eqnarray}
O^{\alpha} = \prod_{r \in P_{\alpha}} \hat{\pi}_{r}^{\alpha}.
\label{symorb}
\end{eqnarray}
Here, $\alpha = x,y,z$ and $P_{\alpha}$ may denote
any plane orthogonal to the cubic 
$\hat{e}_{\alpha}$ axis. 

{\bf (ii)} In the $t_{2g}$ compounds 
(e.g., LaTiO$_3$), we have in $H_{orb}$
of Eq.(\ref{orb}) 
\begin{eqnarray}
\hat\pi_{r}^\alpha = 
\frac{1}{2} \sigma_{r}^\alpha.
\label{compass1}
\end{eqnarray} 
This is called the \emph{orbital compass} model.  
The symmetries of this Hamiltonian are given 
by Eqs.(\ref{symorb}, \ref{compass1}). 
Rotations of individual lower-dimensional planes 
about an axis orthogonal to them
leave the system invariant. The two dimensional orbital compass model
is given by Eqs.(\ref{orb},\ref{compass1}) on the square lattice 
with $\alpha \in \{ x,z\}$ and displays $d=1$ $Z_{2}$ symmetries
(wherein the planes $P$ of Eq.(\ref{symorb}) become lines). 

\newcounter{obrazek}

\begin{figure}[htb]
\vspace*{-0.5cm}
\includegraphics[angle=-90,width=8cm]{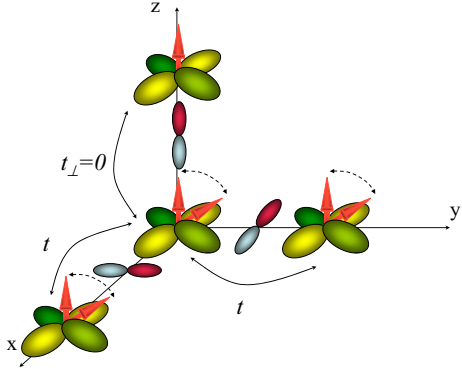}
\vspace*{-0.0cm}
\caption{ From Ref.$^{2}$.
The anisotropic hopping amplitudes leading to 
the Kugel-Khomskii (KK) Hamiltonian. 
Similar
to Ref.$^{13}$
the four lobed states denote the $3d$ orbitals
of a transition metal while the intermediate small $p$ orbitals
are oxygen orbital through which the super-exchange process 
occurs. The dark and bright shades 
denote positive and negative regions
of the orbital wave-function. Due to orthogonality with intermediate 
oxygen $p$ states, in 
any orbital state $|\alpha \rangle$ (e.g. $| Z \rangle 
\equiv | d_{xy} \rangle $ above), hopping is 
forbidden between sites separated along the cubic 
$\alpha$ (Z above) axis. The ensuing super-exchange (KK) Hamiltonian
exhibits a $d=2$ $SU(2)$ symmetry
corresponding to a uniform rotation of all
spins whose orbital state is $|\alpha \rangle$ 
in any plane orthogonal to the cubic direction $\alpha$.}
\label{figu}
\end{figure}

{\bf{b}}) {\it Spins in transition metal compounds}-  Following 
\cite{Harris}, we label the three $t_{2g}$ states $|d_{yz} \rangle, 
| d_{xz} \rangle, | d_{xy} \rangle$ 
by $|X \rangle, | Y \rangle,$ 
and $|Z \rangle$. In the $t_{2g}$ compounds, 
hopping is prohibited via intermediate 
oxygen p orbitals between any two electronic
states of orbital flavor $\alpha$ ($\alpha = X, Y$, or $Z$)
along the $\alpha$ axis of the cubic lattice 
(see Fig.\ref{figu}). As a consequence,
as noted in \cite{Harris}, 
a uniform rotation of all spins, whose electronic orbital state
is $|\alpha \rangle$, in 
any given plane ($P$) orthogonal  
to the $\alpha$ axis 
$c^{\dagger}_{i \alpha \sigma} 
= \sum_{\eta} U^{(P)}_{\sigma, \eta}
d^{\dagger}_{i \alpha \eta}$
with $\sigma, \eta$ the spin 
directions, leaves Eq.(\ref{1}) invariant.
The net spin of the electrons
of orbital flavor $|\alpha \rangle$ 
in any plane orthogonal to the cubic $\alpha$ axis is conserved. 
Here, we have $d=2$ $SU(2)$ symmetries
\begin{eqnarray} 
\hat{O}_{P;\alpha} \equiv [\exp(i\vec{S}^{\alpha}_{P} \cdot 
\vec{\theta}^{\alpha}_{P})/\hbar], ~ ~ [H, \hat{O}_{P;\alpha}]=0,
\label{symt2g}
\end{eqnarray}
with $\vec{S}^{\alpha}_{P} = \sum_{i \in P} \vec{S}_{i}^{\alpha}$,
the sum of all the spins $\vec{S}^{i, \alpha}$ in the orbital state
$\alpha$ in any plane $P$ 
orthogonal to the direction $\alpha$ 
(see Fig.\ref{figu}).

{\bf c})  Superconducting arrays:
A Hamiltonian for superconducting $(p+ip)$ grains 
(e.g. of Sr$_{2}$RuO$_{4}$) 
on a square grid, was recently proposed,  
 \cite{Xu03}
\begin{eqnarray}
H = - K \sum_{\Box} \sigma^{z} \sigma^{z} \sigma^{z} \sigma^{z}
- h \sum_{\bf r} \sigma_{\bf r}^{x}.
\label{XM}
\end{eqnarray}
Here, the four spin product is the product of
all spins common to a given plaquette $\Box$.
The spins reside on the vertices on the plaquette
(not on its bonds as gauge fields).
These systems 
have $(d=1$ $Z_{2}$) 
symmetries similar to those
of the two-dimensional orbital compass model.
With $P$ any row or column,
$\hat{O}_{P} = \prod_{\vec{r} \in  P} \sigma^{x}_{\vec{r}}, 
~~[H, \hat{O}_{P}]=0$.

{\bf d}) Other systems: In \cite{BT} \cite{NBNT}
similar symmetries were found in frustrated spin systems.
Ring exchange Bose metals,
in the absence of nearest neighbor boson
hopping, exhibit $d=1$
symmetries. \cite{Arun02}
Continuous {\em sliding symmetries} of 
Hamiltonians (actions) invariant under 
arbitrary deformations along a transverse direction,
\begin{eqnarray}
\phi(x,y) \to 
\phi(x,y) + f(y),
\label{slideq}
\end{eqnarray} 
appear in many systems. 
Amongst others, such systems were discovered
in works on Quantum Hall liquid crystalline phases,
\cite{lawler} \cite{rad} a number of models of lipid 
bilayers with intercalated DNA strands, \cite{ohern}
and sliding Luttinger liquids. \cite{emery2000}

\section{A theorem on dimensional reduction} 

{\it The absolute mean value of any local quantity 
(involving only a finite number of fields)
which is not invariant under a $d$-dimensional symmetry group 
$G$ of the $D$-dimensional Hamiltonian $H$
is equal or smaller than the absolute mean value of the same quantity 
computed for a $d$-dimensional
Hamiltonian ${\bar H}$ which is globally invariant under  $G$ 
and preserves the range 
of the interactions}. \cite{BN} Non invariant means that the 
quantity under 
consideration, $f(\phi_{\bf i})$, has no invariant component:
\begin{equation}
\sum_k f[{\bf g}_{{\bf i}k}(\phi_{\bf i})] = 0.
\label{nonin}
\end{equation}
For a continuous group, this is replaced by
$\int f[{\bf g}_{\bf i}(\phi_{\bf i})] d{\bf g} = 0$.
To determine if spontaneous symmetry
breaking occurs, we compute 
\begin{equation}
\langle f(\phi_{\bf i}) \rangle = lim_{h\rightarrow 0} 
lim_{N \rightarrow \infty} 
\langle f(\phi_{\bf i}) \rangle_{h,N},
\label{limit}
\end{equation}
where $\langle f(\phi_i) \rangle_{h,N}$ is the mean
value of $f(\phi_i)$ computed on finite lattice of 
$N$ sites and in the presence of a symmetry breaking field $h$. 
Since ${\Lambda}= \bigcup_{l} {\cal C}_l$, 
the site ${\bf i}$ belongs at least to one
set ${\cal C}_j$. It is convenient to rename the fields 
in the following way: 
$\phi_{\bf i}=\psi_{\bf i}$ if ${\bf i} \notin {\cal C}_j$ and 
$\phi_{\bf i}=\eta_{\bf i}$ if ${\bf i} \in {\cal C}_j$. The mean value 
$\langle f(\phi_{\bf i}) \rangle_{h,V}$ is given by:
\begin{eqnarray}
\langle f(\phi_{\bf i}) \rangle_{h,N} =
\frac{\sum_{\{ \phi_{\bf i} \}}  f(\phi_{\bf i}) e^{-\beta H(\phi)} 
e^{-\beta h \sum_{{\bf i}} \phi_{\bf i}}}
{\sum_{\{ \phi_{\bf i} \}} e^{-\beta H(\phi)-\beta h \sum_{{\bf i}} 
\phi_{\bf i}}}=
\nonumber \\
\frac{\sum_{\{ \psi_{\bf i} \}} 
z_{\{ \psi \}} e^{-\beta h \sum_{{\bf i} \notin {\cal C}_j} \psi_{\bf i}} 
[\frac {\sum_{\{ \eta_{\bf i} \} } f(\eta_{\bf i})  
e^{-\beta H(\phi)-\beta h \sum_{{\bf i} \in {\cal C}_j}  
\eta_{\bf i}}}{z_{{\{ \psi \}}}}]}
{\sum_{\{ \psi_{\bf i} \}} 
z_{{\{ \psi \}}} e^{-\beta h \sum_{{\bf i} 
\notin {\cal C}_j} \psi_{\bf i} }  } \nonumber
\\ \mbox{with, ~~} z_{{\{ \psi \}}}= \sum_{\{ \eta_{\bf i} \}} 
e^{-\beta H(\psi,\eta)-\beta h \sum_{{\bf i} 
\in {\cal C}_j}  \eta_{\bf i}}.
\label{mast}
\end{eqnarray}
From Eq.(\ref{mast}):
\begin{equation}
|\langle f(\phi_{\bf i}) \rangle_{h,N}| \leq \mid   
\frac {\sum_{\{ \eta_{\bf i} \} } f(\eta_{\bf i})  
e^{-\beta H({\bar \psi},\eta)-\beta h \sum_{{\bf i} 
\in {\cal C}_j}  \eta_{\bf i}}}{z_{{\{ {\bar \psi} \}}}} \mid, 
\label{final}
\end{equation}
where $\{ {\bar \psi} \}$ is the particular configuration 
of fields ${\psi_{\bf i}}$ that maximizes the 
expression between brackets in Eq.(\ref{mast}). 
$H({\bar \psi},\eta)$ is a $d$-dimensional Hamiltonian for 
the field variables ${\eta}$ which is invariant under the 
{\it global} symmetry group $G_j$ of transformations
${\bf U}_{jk}$ over the field $\eta$. 
We can define ${\bar H}(\eta) \equiv H({\bar \psi},\eta)$. 
The range of the interactions between the 
$\eta$-fields in ${\bar H}(\eta)$ is clearly the same 
as the range of the interactions between the 
$\phi$-fields in $H(\phi)$. This completes the demonstration 
of our theorem.  Note that the ``frozen''
variables ${\bar \psi}_{\bf i}$ act like external fields 
in ${\bar H}(\eta)$ which do not break the global
symmetry group of transformations ${\bf U}_{jk}$. 

{\it Corollary I: Elitzur's theorem.} \cite{Elitzur} Any local quantity 
(i.e. involving only a finite number of fields)
which is not invariant under a local (or $d=0$)
symmetry group has a vanishing mean value at 
any finite temperature. This is a direct consequence of 
Eq.(\ref{mast}) and the fact that ${\bar H}(\eta)$ 
is a zero-dimensional Hamiltonian. \cite{BN}

{\it Corollary II.} \cite{BN} A local quantity which is not gauge 
invariant under a one-dimensional intermediate symmetry group
has a vanishing mean value at any finite temperature for 
systems with finite range interactions. This 
is a consequence of Eq.(\ref{final}) and the absence of spontaneous symmetry
breaking in one-dimensional Hamiltonians
such as
${\bar H}(\eta) \equiv H({\bar \psi},\eta)$ with
interactions of finite range and strength. Here,
$f(\eta_{\bf i})$ is a non-invariant under 
the global symmetry group $G_j$ [see Eq.(\ref{nonin})]. 

{\it Corollary III.} \cite{BN}  In 
finite range systems,
local quantities not invariant 
under continuous two-dimensional symmetries 
have a vanishing mean value at any finite temperature. This results 
from [Eq.(\ref{final})] 
with the Mermin-Wagner theorem \cite{Mermin}:
\begin{equation}
lim_{h\rightarrow 0} lim_{N \rightarrow \infty} 
\frac {\sum_{\{ \eta_{\bf i} \} } f(\eta_{\bf i})  
e^{-\beta H({\bar \psi},\eta)-\beta h \sum_{{\bf i} 
\in {\cal C}_j}  \eta_{\bf i}}}{z_{{\{ {\bar \psi} \}}}} = 0.
\end{equation}
We invoked that $G_j$ is a continuous symmetry 
group of ${\bar H}(\eta)= H({\bar \psi},\eta)$, 
$f(\eta_{\bf i})$ is a non-invariant quantity for 
$G_j$ [see Eq.(\ref{nonin})], and ${\bar H}(\eta)$ is a 
two-dimensional Hamiltonian that only contains finite range 
interactions. 

The generalization of this theorem to the quantum case is 
straightforward if we choose a basis of eigenvectors 
of the local operators linearly coupled to the symmetry 
breaking field $h$. Here, the states can 
be written as a direct product 
$ |\phi \rangle = | \psi \rangle \otimes | \eta \rangle$. 
Eq.(\ref{final}) is re-obtained 
with the sums replaced by traces over the states $| \eta \rangle$:
\begin{equation}
|\langle f({\phi}_{\bf i}) \rangle_{h,N}| \leq    
\frac {{\rm Tr}_{\{ \eta_{\bf i} \} } f( \eta_{\bf i})  
e^{-\beta H({\bar \psi},{ \eta})-\beta h \sum_{{\bf i} 
\in {\cal C}_j}  { \eta}_{\bf i}}}
{{\rm Tr}_{\{ \eta_{\bf i} \} }  
e^{-\beta H({\bar \psi},{\eta})-\beta h \sum_{{\bf i} 
\in {\cal C}_j}  {\eta}_{\bf i}}}, 
\label{quantum}
\end{equation}
In this case, $| {\bar \psi} \rangle$ corresponds to one particular 
state of the basis $| \psi \rangle$ that maximizes
the right side of Eq.(\ref{quantum}). Generalizing standard proofs,
e.g. \cite{assa}, we find 
a  zero temperature quantum extension of 
Corollary III in the presence of a gap:

{\it Corollary IV.} \cite{BN} If a gap exists
in a system possessing a $d \le 2$ dimensional continuous symmetry
in its low energy sector then the expectation value of 
any local quantity not invariant under this symmetry, 
strictly vanishes at zero temperature. Though local
order cannot appear, multi-particle (incl. topological)
order can exist. 

{\it Corollary V.}  
The absolute values of non-symmetry invariant correlators $|G| \equiv 
|\langle \prod_{i \in \Omega_{j}} \phi_{i} \rangle|$
with  $\Omega_{j} \subset C_{j}$ are bounded
from above by absolute values of the same correlators $|G|$ 
in a $d$ dimensional system defined by $C_{j}$ in the
presence of transverse non-symmetry breaking fields.
If no resonant terms appear in the lower dimensional 
spectral functions (due to fractionalization), 
this allows for fractionalization of non-symmetry 
invariant quantities
in the higher dimensional system.

\section{ Consequences of the theorem}

({\bf a})  {\em Spin nematic order in $t_{2g}$ systems}: If 
the KK Hamiltonian (Eq.(\ref{1})) captures the
spin physics of $t_{2g}$ compounds,  then
no magnetization can 
exist at finite temperature\cite{BN} due to the 
continuous $d=2$ symmetries \cite{Harris} that it 
displays (Eq.(\ref{symt2g})). \cite{BN} \cite{Harris} Empirically, 
low temperature magnetization is detected.
Thus, the KK Hamiltonian
of Eq.(\ref{1}) may be augmented by other interactions
which lift this symmetry. 
The simplest quantities invariant
under these symmetries are nematic order parameters.
In the presence of orbital ordering in the $| \alpha \rangle$ state,
superpositions of $\langle \vec{S}_{r} \cdot 
\vec{S}_{r+ n \hat{e}_{\eta}} \rangle$, with $\eta = x,y,z$ where 
$\eta \neq \alpha$ and $n$ an integer, 
need not vanish. If the KK Hamiltonian embodies
the dominant contribution to the spin physics, nematic order
might persist to far higher temperatures than
the currently measured magnetization. \cite{BN}

({\bf b}) {\em Orbital order}: 
The orbital only Hamiltonians discussed
earlier exhibit a $d=2$ {\em discrete} $Z_{2}$ 
symmetry. The theorem \cite{BN} allows such symmetries to be
broken. Indeed, as we will review shortly, in these 
orbital only Hamiltonians, order already appears
at the classical level- a tendency which may be 
enhanced by quantum
fluctuations.

({\bf c}) {\em Nematic orbital order in two dimensional 
(p+ip) superconducting arrays and two dimensional
orbital systems}: The two dimensional $(p+ip)$ superconducting 
arrays of Eq.(\ref{XM}) exhibit a $d=1$ 
$Z_{2}$ symmetry. As these symmetries
cannot be broken, no magnetization 
can arise, $\langle \sigma_{\alpha} \rangle =0$.
The simplest symmetry allowed order 
parameter is of the nematic type
which is indeed realized classicaly. \cite{NBCv} \cite{BCN} \cite{Mishra}

({\bf d}) {\em Fractionalization in spin and orbital systems}:
Corollary (V) allows for fractionalization
in quantum systems where d= 1, 2. It enables
symmetry invariant quasi-particles excitations to {\em coexist}
with non-symmetry invariant fractionalized excitations. 
Fractionalized excitations 
may propagate in $d_{s} = D- d$ dimensional regions
(with $D$ the spatial dimensionality of the system). 
Examples 
afforded by several frustrated spin models 
where spinons may drift along
lines ($d_{s}=1$) on the square lattice \cite{BT}
and in $d_{s}=D$ dimensional regions on the
pyrochlore lattice. \cite{NBNT} 

({\bf e}) {\em Absence of charge order}:
In systems, such as quantum Hall smectics, in which the 
system is invariant to the charge density variations 
of Eq.(\ref{slideq}), we have $\langle \phi \rangle =0$.

\section{Order by disorder in symmetry allowed instances}

When symmetry breaking is allowed 
(e.g. the two dimensional Ising symmetry (Eq.(\ref{symorb})) 
of the 120 $^\circ$ Hamiltonian), 
order often transpires
by a fluctuation driven 
mechanism (``order by disorder'').
\cite{Shender&Henley} 
Although several states may
appear to be equally valid candidate ground state,
fluctuations can stabilize those states which have
the largest phase space volume 
for low energy fluctuations about them. 
These differences are captured in values of
the free energies for fluctuations about the contending
states. Classicaly, fluctuations are driven by 
thermal effects. Quantum   
tunneling processes may fortify such tendencies. 


If the Pauli matrices
$\sigma$ in Eq.(\ref{orb}) are replaced by the spin $S$ 
generators and the limit $ S \to \infty$ is taken then we will obtain the 
classical 120$^\circ$ model. Here, the 
{\em free energy} has strict
minima for six uniform orientations \cite{NBCv} \cite{BCN}
$\vec{S}_{i} = \pm S \hat{a}, \vec{S}_{i} = \pm S \hat{b}, \vec{S}_{i} = \pm 
S \hat{c}$.
Out of the exponentially large number of ground states 
(supplanted by an additional global $U(1)$ rotational symmetry
which emerges in the ground state sector),
only six are chosen. Interfaces between 
uniform states (such as that borne
by the application of d=2 $Z_{2}$ reflections on a 
uniform state, see fig.(\ref{figrubick})) leads to a surface tension additive 
in the number of symmetry operations. Being of an entropic
origin, the surface tension  
between various uniform domains
is temperature independent
and does not diverge at low temperatures. \cite{NBCv} \cite{BCN} 
Orbital order already appears
within the classical (formally, $S \to \infty$)
limit \cite{NBCv} \cite{BCN} and is not exclusively reliant on  
subtle quantum zero 
point fluctuations (captured by  $1/S$ calculations) 
for its stabilization. 
Indeed, orbital order is detected 
up to relatively high temperatures
$({\cal{O}}(100 K)$). \cite{RXRays}

\section{Dualities as space-time reflections}

\begin{figure}
\centerline{\psfig{figure=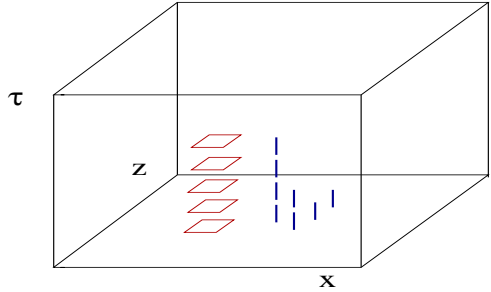,height=3.81cm,width=6.4cm,angle=0}}
\caption{From Ref.$^{5}$. The classical Euclidean action corresponding
to the Hamiltonian of Eq.(\ref{XM}) at zero temperature
in a basis quantized along the $\sigma^{z}$ direction. The transverse
field leads to bonds parallel
to the imaginary time axis while
the plaquette interactions become
replicated along the imaginary time axis.
Taking an equal time slice of this system,
we find the four spin term of Eq.(\ref{XM})
and the on-site magnetic field
term. If we interchange $\tau$ with $z$,
we find the planar orbital compass model 
in the basis quantized along the $\sigma^{x}$ direction.}
\label{FIG:DUAL1}
\end{figure}

Explicit operator representations
show that the two dimensional variant of the orbital compass model 
(Eqs.(\ref{orb}, \ref{compass1}) is 
dual to the Xu-Moore model of $(p+ip)$ 
superconducting arrays (Eq.(\ref{XM}). \cite{NF} We now 
examine this 
duality in the discrete Euclidean path 
integral formulation. This examination illustrates
how geometrical reflections may lead 
to dualities. \cite{NF} In a basis quantized along 
$\sigma^{z}$, the zero temperature Euclidean action of 
the two dimensional orbital compass model is 
\begin{eqnarray}
S =&&\!\!\!\! - K_{x} \sum_{\Box \in (x \tau) ~ \mbox{plane}} 
\sigma^{z}_{r,\tau} 
\sigma^{z}_{r, \tau + \Delta \tau} \sigma^{z}_{r + \hat{e}_{x}, \tau}
\sigma^{z}_{r + \hat{e}_{x}, \tau + \Delta \tau} 
 -
(\Delta \tau) J_{z} \sum_{r} \sigma^{z}_{r} 
\sigma^{z}_{r+e_{z}}.
\end{eqnarray}
A schematic of this action in Euclidean space-time
is shown in Fig.(\ref{FIG:DUAL1}).
If we relabel the axes and replace the spatial index $x$ with the
temporal index $\tau$, we will immediately find the classical action 
corresponding to the the Hamiltonian of Eq.(\ref{XM}) depicting
$p+ip$ superconducting grains in a square grid. This 
suggests that the planar orbital compass system 
and the $(p+ip)$ Hamiltonian (Eq.(\ref{XM})) 
are dual to each other as indeed occurs at
all temperatures. 
Strong-weak coupling dualities 
that these Hamiltonians (and others) display 
can be similarly established. \cite{NF} 

This work was supported, in part, by the DOE at LANL (CDB), 
and by the National Science Foundation through grants 
NSF DMR 0442537 at UIUC (EF).

\end{document}